\documentclass[9pt,twocolumn,twoside]{pnas-new}

\templatetype{pnasresearcharticle} 

\title{Extent of Fermi-surface reconstruction in the high-temperature superconductor HgBa$_2$CuO$_{4+\delta}$}
\author[a,1]{Mun K. ~Chan}
\author[a]{Ross D. McDonald} 
\author[d]{Brad J. Ramshaw}
\author[a]{Jon B. Betts}
\author[b]{ Arkady Shekhter}
\author[c]{Eric D. Bauer}
\author[a] {Neil Harrison}

\affil[a]{Pulsed Field Facility, National High Magnetic Field Laboratory,~Los Alamos National Laboratory,~Los Alamos,~New Mexico 87545, USA}
\affil[b]{National High Magnetic Field Laboratory, Florida State University, Tallahassee, Forida 32310, USA}
\affil[c]{Los Alamos National Laboratory, Los Alamos, New Mexico, 87545, USA}
\affil[d]{Laboratory of Atomic and Solid State Physics, Cornell University, Ithaca, NY 148.3, USA}

\leadauthor{Chan} 

\correspondingauthor{\textsuperscript{1}To whom correspondence should be addressed. E-mail: mkchan@lanl.gov}

\begin{abstract}

High magnetic fields have revealed a surprisingly small Fermi-surface in underdoped cuprates, possibly resulting from Fermi-surface reconstruction due to an order parameter that breaks translational symmetry of the crystal lattice. A crucial issue concerns the doping extent of this state and its relationship to the principal pseudogap and superconducting phases. We employ pulsed magnetic field measurements on the cuprate HgBa$_2$CuO$_{4+\delta}$ to identify signatures of Fermi surface reconstruction from a sign change of the Hall effect and a peak in the temperature-dependent planar resistivity. We  trace the termination of Fermi-surface reconstruction to two hole concentrations where the superconducting upper critical fields are found to be enhanced. One of these points is associated with the pseudogap end-point near optimal doping. These results connect the Fermi-surface reconstruction to both superconductivity and the pseudogap phenomena.  
\end{abstract}

\dates{This manuscript was compiled on \today}
\begin{document}
\maketitle
\thispagestyle{firststyle}
\ifthenelse{\boolean{shortarticle}}{\ifthenelse{\boolean{singlecolumn}}{\abscontentformatted}{\abscontent}}{}


\dropcap{A} broad paradigm of unconventional superconductivity stipulates that it results from quantum critical fluctuations associated with a continuous phase boundary that ends at a zero temperature quantum critical point~\cite{natphysQCP}. For the superconducting cuprates, the termination of the pseudogap phase at a critical doping $p^*$ located near optimal doping, has long been speculated to be a quantum critical point~\cite{keimer15,varma97,kivelson98}. It is thus important to identify clear signatures of broken symmetry states and ascertain their relationship to the pseudogap and to possible quantum critical points. Large magnetic fields, by suppressing the superconductivity in underdoped YBa$_2$Cu$_3$O$_{6+x}$ (Y123) and Hg1201, have revealed a sign-change of the Hall effect at a characteristic temperature $T_{\rm H}$~\cite{leboeuf11,doiron13}. Quantum oscillations measurements at low temperature also indicates the presence of a small Fermi-pocket~\cite{leyraud07,sebastian08,barisic13b,chan16b}. These results have been interpreted as a consequence of Fermi-surface reconstruction by an order parameter that breaks the translational symmetry of the crystal lattice~\cite{leboeuf11,doiron13,millis07,harrison11,chakravarty08}. Here, we present a doping-dependent study of Fermi-surface reconstruction in Hg1201, which features a relatively simple tetragonal crystal structure comprising one CuO$_2$ plane per unit cell~\cite{putilin93}. Its high maximal superconducting temperature necessitates large pulsed magnetic fields, up to 90 tesla employed in this work, to suppress the superconductivity at low temperatures (see Materials and Methods). 
 

\begin{figure*}[t] 
\includegraphics*[width=.95\textwidth]{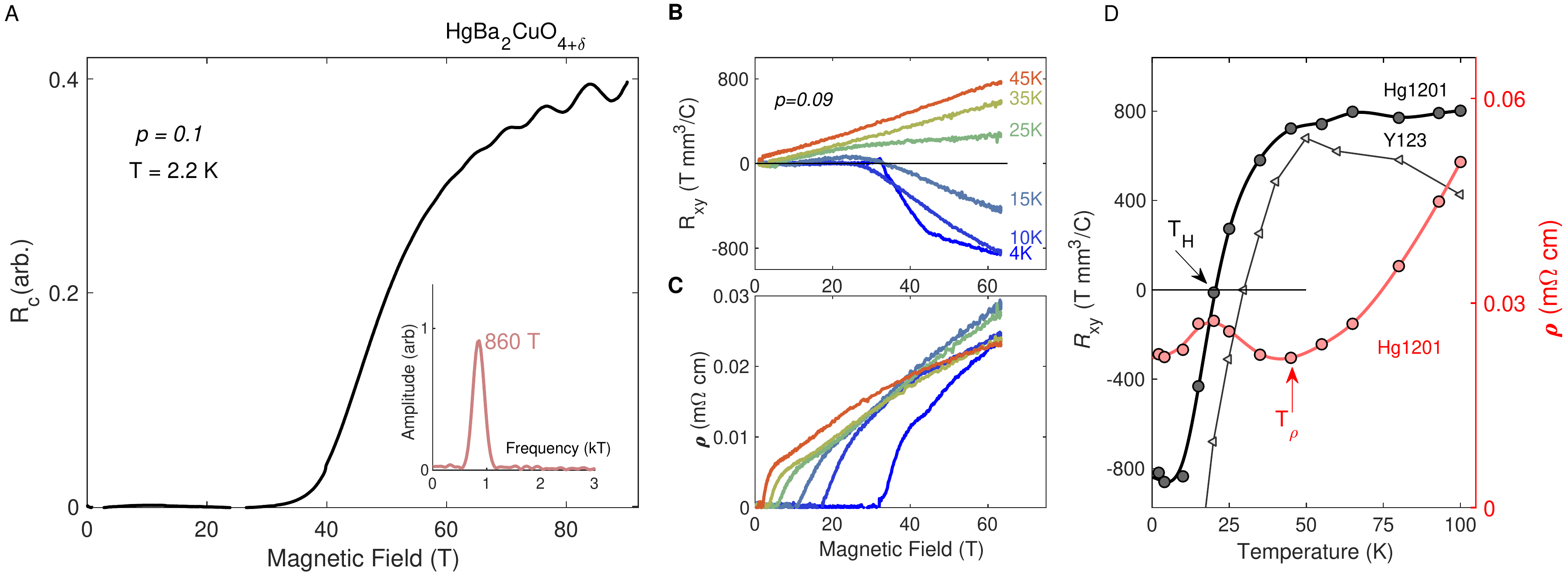}%
\caption{{\bf Signatures of Fermi surface reconstruction in Hg1201} {\bf A,} Magnetic field dependence of c-axes resistance for Hg1201 with doping $p = 0.1$ and $T_c = 71$~K. Quantum oscillations are observed at high fields. Inset: The quantum oscillation spectrum, from a Fourier-transform of the data between 45 T and 90 T after subtracting a background, comprises a single peak at 860~T. {\bf B\&C} Pulse-magnetic field measurements of the Hall resistance $R_{\rm xy}$ and transverse planar resistivity $\rho$ at different temperatures for sample UD69, $T_c = 69$~K, corresponding to $p=0.09$. {\bf D,} The extracted temperature dependence at 64 T of $\rho$ (red) and $R_{\rm xy}$ (black) from panels {\bf B\&C}. Characteristic temperatures marked by the upturn in $\rho$  and  sign-change of $R_{\rm H}$ are indicated by arrows. The sign change in $R_{\rm xy}$ is consistent with that observed in Y123 $p=0.10$ measured at 61 T(triangles, reproduced from Ref.~\cite{leboeuf07}). 
}
\label{QO}
\end{figure*}

\begin{figure*}[t] 
 \includegraphics*[width=.92\textwidth]{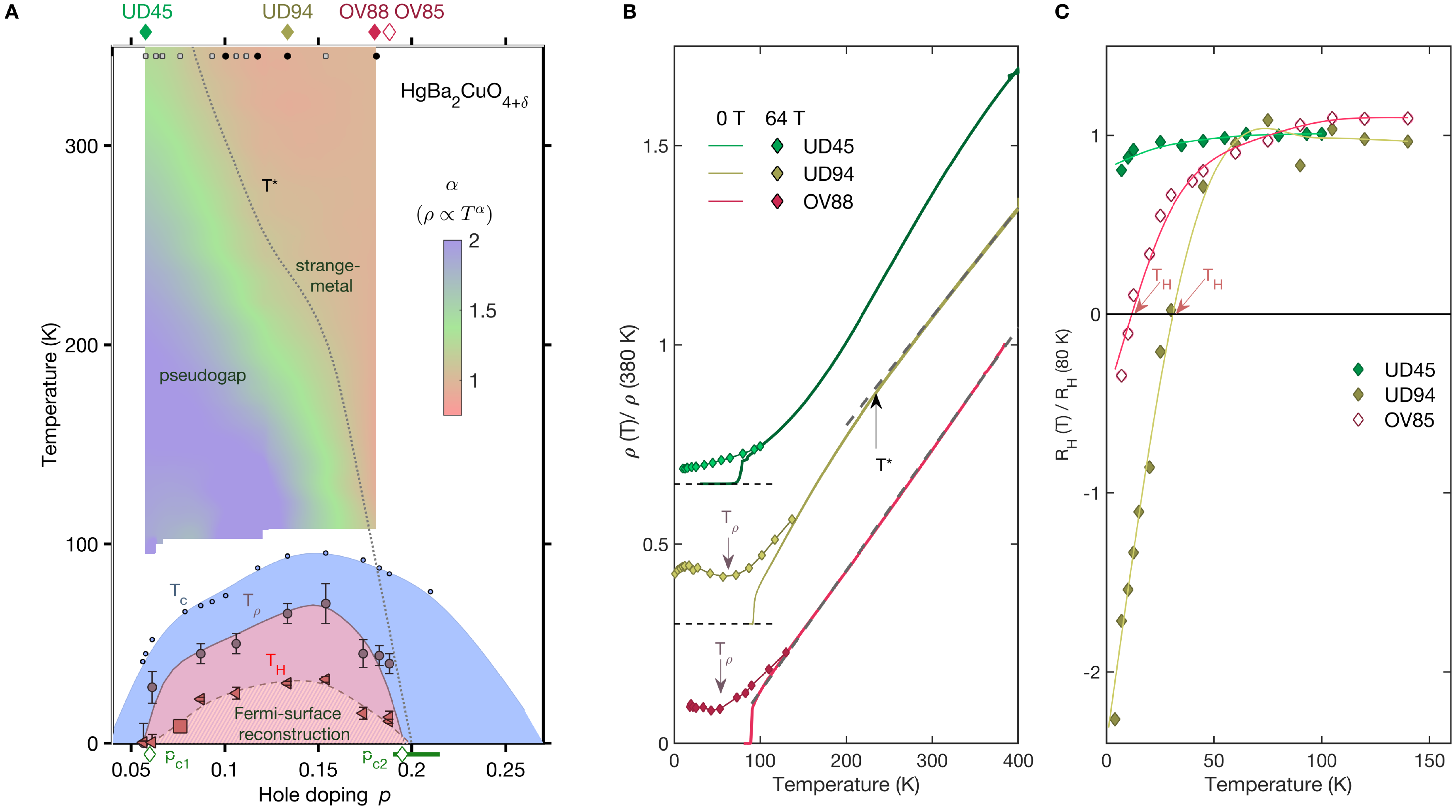}%
\caption{{\bf Characteristic temperatures in Hg1201 from electrical transport.} {\bf A,} Doping dependence of characteristic temperatures for Hg1201. The color contour represents the temperature exponent  $\alpha$ of the zero-field planar resistivity of the form $\rho=\rho_0+A T^\alpha$, where $A$ is a temperature independent coefficient. Black and grey squares along the top edge of the panel mark doping levels measured in this work and in Ref.~\cite{barisic15}, respectively, used to create the contour. $T^*$ marks the deviation from strange-metal behavior ($\rho\propto T^1$), which is represented here by the $\alpha =1.1$ contour line (dashed black line). This determination of $T^*$ is consistent with the the onset of magnetic\cite{li08,tang18} and nematic~\cite{murayama19}  orders (see SI Appendix Fig. S13). $T_\rho$ and $T_{\rm H}$ are characteristic temperatures determined from the planar resistivity and the Hall coefficient respectively. The square symbol is $T_{\rm H}$ from Ref.~\cite{doiron13}. {\bf B,} Planar resistivity $\rho$  in zero field (solid lines) and at 64 T (symbols) for representative doping levels indicated on top of panel {\bf A}. The data are shifted vertically for clarity. $T^*$ (deviation from $\rho\propto T$) and $T_{\rho}$ are indicated by arrows. The dashed black lines are linear fits indicating strange metal behavior. {\bf C,} Temperature dependence of Hall coefficient $R_{\rm H}(T)$. The change in sign occurs at $T_{\rm H}$, indicated by the arrows.
}
\label{PDdoping}
\end{figure*}

\begin{figure}[t] 
\centering
 \includegraphics*[width=.95\linewidth]{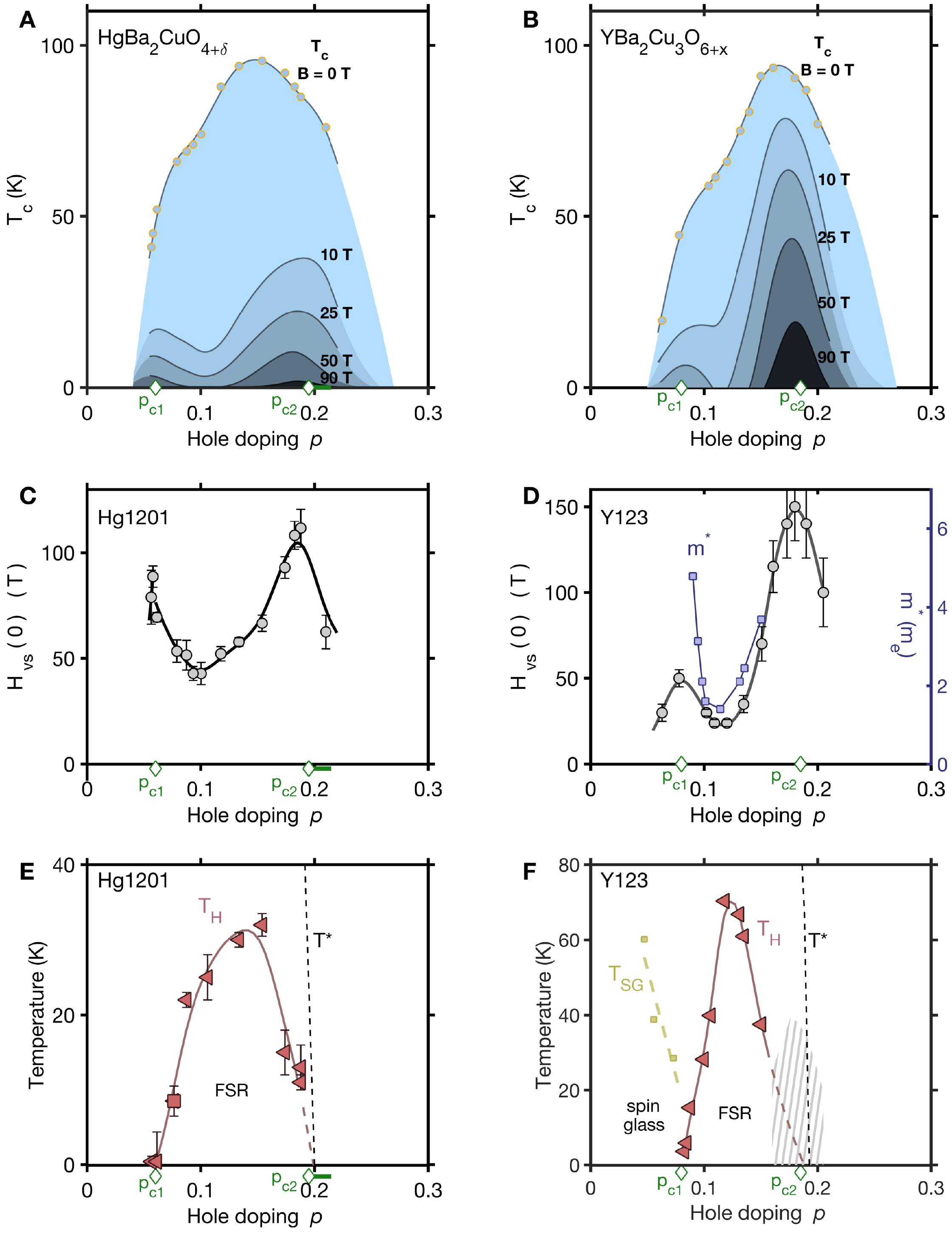}%
\caption{{\bf Superconductivity and the boundaries of Fermi-surface reconstruction} 
  {\bf A,} Magnetic field dependence of $T_c$  for Hg1201, defined as the temperature below which the resistivity is zero measured for samples with zero-field $T_c$'s indicated by yellow circles. {\bf B,} $T_c$ for Y123 from Ref.~\cite{griss14}.  {\bf C,} Vortex-melting field at zero temperature $H_{\rm vs}(0)$ for Hg1201 (see Methods and Materials for determination of these values). {\bf D,} $H_{\rm vs}(0)$ for Y123 from Ref.~\cite{griss14}.  Effective mass $m^*$ from quantum oscillations reported in Refs.~\cite{sebastian10c,ramshaw15} (purple squares, right axes).  {\bf E}, Characteristic temperature corresponding to the sign change of the Hall coefficient $T_{\rm H}$ in Hg1201. The pseudogap temperature $T^*$ (dashed line) is an extrapolation from high temperature transport data, see Fig.~\ref{PDdoping}A. {\bf F,} $T_H$ for Y123 from Refs.~\cite{leboeuf07,leboeuf13,badoux16}. $R_{\rm H}$ in Y123 has not been reported within the hatched grey region around $p_{\rm c2}$. A glassy spin-density wave phase appears below $T_{\rm SG}$ in the far underdoped regime~\cite{haug10}, that is apparently absent in Hg1201 (see SI Appendix Fig.~S2).  
}
\label{PDTc}
\end{figure}

\subsection*{Signatures of Fermi-surface reconstruction from electrical transport}
Fig.~\ref{QO}A shows the {\it c}-axes resistance for an underdoped Hg1201 crystal exhibiting clear quantum oscillations with a frequency of 860~T that corresponds to a small pocket comprising $\approx 3\%$ of the Brillouin zone. The Hall resistance $R_{xy}$, Figs.~\ref{QO}B~\& D,  undergoes a downturn and eventual sign change at $T_{\rm H}\approx 20~$K. These results attest to the high-quality of the present crystals and confirm prior results indicating Fermi-surface reconstruction in Hg1201 at similar doping levels~\cite{doiron13,barisic13b,chan16b}. 

We have also found a peak in the planar resistivity $\rho (T)$ at high fields, Figs.~\ref{QO}C\&D. $\rho (T)$ upturns at a temperature coincident with the steep downturn in $R_{xy} (T)$, as shown in Fig.~\ref{QO}D and SI Appendix Fig.~S7. The common characteristic temperatures suggest the features in $\rho(T)$ are related to the sign-change of the Hall effect, and thus Fermi-surface reconstruction. The behavior of $\rho(T)$ is reminiscent to that associated with phase transitions that open a gap on the Fermi-surface such as charge- or spin- density waves~\cite{fawcett88SDW,gruner17,morosan06,ong77}. In these systems, the initial increase in $\rho(T)$ is interpreted as a reduction in charge carriers by a gap opening. Upon decreasing temperature further, $\rho (T)$ peaks as the gap becomes fully formed and metallic behavior resumes. For Hg1201, we are able to identify the temperature at which $\rho(T)$ upturns over a broad doping range (SI Appendix Fig.~S7), which we label $T_{\rho}$. Although $T_{\rho}$ might signify the temperature at which a gap opens, the overall peak feature in Hg1201 is  relatively broad, and lacks the distinct kink associated with typical density-wave phase transition~\cite{fawcett88SDW,gruner17,morosan06,ong77} (see SI Appendix Fig.~S11). The shape of the broad peak likely results from a complicated interplay between the effects of disorder on the gap size, and the effects of scattering rate on the temperature and magnetic field dependencies of $\rho$.  


\subsection*{Fermi-surface reconstruction boundary}
Fig.~\ref{PDdoping}A shows $T_\rho$ and $T_{\rm H}$ determined from measurements of samples over a wide doping range (see SI Appendix Figs.~S3,~S5-S7 for underlying field and temperature dependent data). We note that $T_{\rm H}$ is determined at sufficiently high-magnetic fields such that it is field independent (see SI Appendix Figs.~S3~\&~S4)~\cite{leboeuf07}. The larger error bars on our determination of $T_\rho$ reflect the broad minimum in $\rho (T)$. The doping dependence of both $T_{\rm H}$ and $T_\rho$  forms domes that peak at optimal doping and approaches zero temperature at $p_{\rm c1} \approx  0.060 (\pm 0.005)$ near the underdoped edge of the superconducting dome and at $p_{\rm c2}\approx  0.195 (-0.005,+0.02)$ which is slightly overdoped.

\subsection*{Superconductivity in applied magnetic field}
In zero magnetic field, Hg1201 features a superconducting dome with a broad maximum at $p\approx 0.16$ (Fig.~\ref{PDTc}A).  In a magnetic field, $T_c$, defined by the appearance of zero resistivity, is suppressed unequally across the dome, revealing two islands of superconductivity.  This is a reflection of the doping evolution of the vortex solid melting field $H_{\rm vs}$. We determine $H_{\rm vs}(T\rightarrow 0)$ by fitting the melting line measured at finite temperatures to a mean-field Ginzburg-Landau treatment (see Materials and Methods)~\cite{houghton89}. In this treatment, which only accounts for vortex melting via thermal fluctuations, the upper critical field is equivalent to the vortex melting field at zero temperature $H_{c2}^{MF}(0) = H_{\rm vs}(0)$.  $H_{\rm c2}^{\rm MF}$ for Y123~\cite{ramshaw12,griss14,Zhou13148NMR} and Tl$_2$Ba$_2$CuO$_{6+\delta}$~\cite{griss14} were reported to be consistent with direct measures of $H_{c2}$. However, it has been argued that $H_{c2}^{MF}$ is lower than the actual upper critical field~\cite{yu16}. Being unable to measure $H_{\rm c2}$ at low temperatures directly, we take $H_{\rm c2}^{\rm MF}=H_{\rm vs} (0)$ as a lower limit on $H_{\rm c2}$. The fitted doping dependence of $H_{\rm vs} (0)$ is non-monotonic and is largest upon approaching the critical dopings of Fermi-surface reconstruction at $p_{\rm c1}$ and $p_{\rm c2}$, as shown in Fig.~\ref{PDTc}C. As a result each of the two critical points underlie a separate dome of superconductivity in high-magnetic fields.  

\subsection*{Pseudogap temperature from electrical transport}
The pseudogap temperature $T^*$ is typically identified in electrical transport as the deviation from high-temperature linear-in-temperature resistivity ($\rho\propto T$) termed strange-metal behavior. For Hg1201, we combine prior~\cite{barisic19} and new measurements of the planar resistivity in zero-magnetic field to determine the doping dependence of $T^*$, which is consistent other markers of $T^*$ as shown in SI Appendix Fig.~S13. We find $T^*$ decreases with increasing hole doping and intersects the superconducting dome at $p\sim 0.18$ (see OV88 resistivity in Fig.~\ref{PDdoping}B). The superconductivity prevents the observation of $\rho\propto T$ to zero temperature without the application of a large magnetic field, which itself modifies the temperature dependent behavior~\cite{gallo18}. Nevertheless,  $T^*$ extrapolates from temperatures above $T_c$ to zero temperature at doping $p^*\approx p_{\rm c2}=0.195$.  This is consistent with the pseudogap critical point of $p^*\sim 0.19$ determined from spectroscopic probes in the cuprates~\cite{vishik12,fujita14,tallon01}. Our results show that the phase responsible for the Fermi-surface reconstruction likely terminates near the pseudogap endpoint. 

\subsection*{Discussion}
It is instructive to compare Hg1201 to Y123, which has been extensively studied at high-magnetic fields, and is the only other underdoped cuprate synthesized with sufficient quality to yield quantum oscillations at low temperatures.  As shown in Fig.~\ref{PDTc}, the doping evolution of $T_c$, $H_{\rm vs}(0)$, and $T_{\rm H}$  in Y123 and Hg1201 are very similar, thus indicating generic features of the cuprates. A crucial difference between the two is the doping extent of $T_{\rm H}$: it was reported that the sign change of the Hall effect in Y123, and thus the Fermi-surface reconstruction, exists only in a limited underdoped regime  and terminates below $p \approx 16$, which is clearly seperated from the pseudogap termination at $p^*\approx 0.19$~\cite{badoux16}. For Hg1201, we find that $T_{\rm H}$ and $T_\rho$ remains finite at optimal doping and extends into the overdoped regime. Factors limiting the apparent doping extent of finite $T_{\rm H}$ in the overdoped regime in Y123 are discussed in SI Appendix note 1. Our determination of $p_{c2}$ has a relatively large uncertainty compared to $p_{\rm c1}$, as it is based on extrapolating measurements on samples with $p<p^*$. Measurements on other cuprates suggest that the Fermi-surface must transition into a large hole-like Fermi-surface in the very overdoped regime $p\sim0.3$~\cite{vignolle08,hussey03}. How this process occurs in Hg1201 cannot presently be answered.

Phenomenological treatments of Fermi-surface reconstruction by various order parameters have been proposed to explain the high-magnetic field data in underdoped Hg1201 and Y123~\cite{millis07,harrison11,chakravarty08,briffa15,maharaj15}. For Y123, extensive high-field measurements have found both a unidirectional and bidirectional CDW~\cite{gerber15,chang16,wu13}. The correlation length, $\xi_{\rm CDW}$, approaches a few hundred Angstroms~\cite{gerber15,chang16} in high-magnetic fields and thermodynamic signatures of a second order phase transition~\cite{leboeuf13,laliberte18} have also been observed. It is thus likely that either a unidirectional~\cite{maharaj15,gannot19} and/or bidirectional~\cite{sebastian14,briffa15,laliberte18} CDW is responsible for Fermi-surface reconstruction in Y123.

On the other hand, CDW correlations in Hg1201, measured only down to temperatures near  $T_c$, have a much shorter $\xi_{\rm CDW}\sim20-30 ~\AA$~\cite{tabis14,tabis17} than in Y123. It is unclear whether a short correlation length CDW can yield the observed reconstructed Fermi-surface. A theoretical lower bound for $\xi_{\rm CDW}$ has been proposed~\cite{gannot19} based on an effective Dingle damping-factor of the quantum oscillation amplitude: $R_D = e^{-B_D/B}$, where $B_D\gtrsim 2n\hbar k_F/e\xi_{\rm CDW}$, and $n=1~{\rm or}~ 2$ for unidirectional and bidirectional CDW reconstruction respectively. The Fermi-momentum is given by $k_F = \sqrt{2 e F/\hbar}$ where $F$ is the quantum oscillation frequency. The quantum oscillations in Fig.~\ref{QO}a (see SI Appendix Fig.~S12) yields $B_D^{exp.} = 450~$T, and thus a lower bound of  $\xi_{\rm CDW}\gtrsim45~\AA~{\rm or}~90~\AA$ for unidirectional and bidirectional CDW reconstruction respectively. Although such an analysis would imply that unidirectional CDW is more compatible with the small  $\xi_{\rm CDW}$, as was also claimed for Y123~\cite{gannot19}, the size of the Fermi-pocket determined from quantum oscillations in Hg1201 is more consistent with bidirectional CDW reconstruction~\cite{chan16b,tabis14}. There is presently no direct measures of the $\xi_{\rm CDW}$ in Hg1201 in high-fields and low temperatures to determine whether $\xi_{\rm CDW}$ increases significantly. Considering the similarity of the high-field electrical transport between Y123 and Hg1201, it is tempting to assume that CDW is generally responsible for the Fermi-surface reconstruction in both materials. However, the case for CDW induced reconstruction in Hg1201 is presently weak. NMR up to 30 T in Hg1201  found evidence of a CDW, but no indication that its amplitude or correlation length increases with magnetic field~\cite{lee2017}. Notably, 30 T is smaller than then fields required to observe Fermi-surface reconstruction in the present work.

Regardless of the exact cause of reconstruction, we have determined that its characteristic temperatures approach zero at both conjectured quantum critical points where $H_{\rm vs}(0)$ is enhanced, including near the pseudogap endpoint at $p^*$. For Y123, the presence of two quantum critical points is supported by measurements of doping dependent effective electronic mass $m^\star$ from quantum oscillations~\cite{ramshaw15,sebastian10c}, and from indirect measures of effective mass such as $H_{\rm vs}(0)$ (Fig.~\ref{PDTc}D~\cite{griss14} and the specific heat jump at $T_{\rm c}$~\cite{ramshaw15,loram93}. 

 Although CDW has been theoretically suggested to be quantum critical~\cite{caprara17,castellani95}, and is a possible candidate for the cause of Fermi-surface reconstruction, it is worth noting that the pseudogap state and the anomalous truncated Fermi arcs which define it~\cite{vishik102}, remain the overarching mystery of the underdoped cuprate electronic structure. The Fermi-surface reconstruction examined here occurs in the backdrop of this enigmatic state. In addition to broken translational symmetry suggested by Fermi-surface reconstruction, broken time-reversal symmetry in the form of intra-unit cell magnetic order~\cite{fauque06,li08,zhao17} and broken-planar-fourfold rotational symmetry termed nematicity~\cite{sato17,murayama19} have been detected near $T^*$ of Y123 and Hg1201. The present determination of the extent of Fermi-surface reconstruction and its connection to the pseudogap and quantum critical points clarifies the high-field phase diagram of the cuprates. It emphasizes the need for theoretical treatments of the pseudogap that encompasses the Fermi-surface reconstruction in addition to the other broken symmetries  {\it e.g.} Refs.~\cite{hayward14,Nie14,chakraborty19,varma19}.



\matmethods{
\subsection*{Sample preparation} Hg1201 single crystals were grown using an encapsulated self-flux method~\cite{yamamoto00, zhao06} at Los Alamos National Laboratory. Quality of the samples is evinced by sharp superconducting transitions (full transition widths of 1-3 ~K depending on doping, see SI Appendix Fig.~S8) and large quantum oscillations (see Fig.~\ref{QO}a). The samples are labelled based on their doping (underdoped (UD) or overdoped (OV)) followed by the temperature corresponding to the mid-point of the superconducting diamagnetic transition. The quoted hole concentrations of all our samples are based on the phenomenological Seebeck coefficient scale~\cite{yamamoto00}.

\subsection*{Pulsed field measurements} High magnetic field measurements were performed
at the Pulsed-Field Facility of the National High Magnetic Field Laboratory at Los Alamos National Laboratory. Two different magnet systems were used: the 65T and 100T multishot magnets. The 65T magnet is capacitor driven with a full width of $\sim 100$~ms. The 100T magnet consists of an inner and outer magnets. The outer magnet is first generator driven relatively slowly ($\sim 3$ s total width) between 0 and 37 T, followed by a faster ($\sim 15$ ms) capacitor bank driven pulse to 90 T. Electrical transport samples, with typical dimensions of $\lesssim 1.5\times0.5\times0.1 {\rm mm}^3$ along a,b, and c directions, had 4 to 6 wires attached to Au pads on freshly cleaved sample edges. Planar Hall and resistivity data were generated by taking the difference and average respectively between magnet pulses along the crystalline $c$ and -$c$ directions. $T_{\rm H}$ was determined with magnetic field pulses up to 90~T for samples UD45, UD94 , and OV85. 65~T field pulses were used for all other samples. $T_\rho$ was determined at high-magnetic fields between 60-65~T.  Pulsed-field magnetization was performed using an extraction magnetometer by taking the difference between sample-in and sample-out pulses. The temperature dependence of $\rho$ in zero magnetic field was taken with a Quantum Design PPMS. 

\subsection*{Vortex-solid melting line, $T_c$, and $H_{\rm c2}$ }
The  $T_c(p)$ curves for different applied fields in Fig.~\ref{PDTc}A are determined from interpolating the vortex-melting curves (SI Appendix Fig.~S10) at each hole concentration studied. The vortex-solid melting field, $H_{\rm vs}$, at each temperature was determined by the onset of non-zero resistance in magnetic field dependent isoterms such as those shown in SI Appendix Figs.~S5~\&~S9. Either four-point {\it ac} electrical transport or contactless resistivity~\cite{chan16b} methods were used.  $H_{\rm vs}$ from magnetization measurements in pulsed (up to 65 tesla) and DC~\cite{eley19} (up to 7 tesla) fields is defined as the field below which the up and down sweeps are hysteretic. The determination of $H_{\rm vs}$ from magnetization measurements on UD95 are consistent with that determined from resistivity, as shown in SI Appendix Fig.~S9 

The measured melting curve at each doping was fit to the following expression for flux lattice melting due to thermal fluctuations based on a mean-field Ginzburg-Landau description~\cite{houghton89}

\begin{equation}
\left[\frac{1}{(1-t)^{1/2}}\right]   \left[\frac{b(t)^{1/2}}{(1-b(t))} \right] \left[\frac{4\sqrt{2}-1}{(1-b(t))^{1/2}}+1\right] = \alpha \equiv \frac{2\pi c_L^2}{\sqrt{G_i}}
\end{equation}
where $t=T/T_c$ and $b(t) = H_{\rm vs}(t) /H_{\rm c2}(t)$. $G_i\propto (k_b T_c)^2\gamma^2\lambda_{ab}^4/\xi_{ab}^2$ is the Ginzburg parameter which depends on the mass anisotropy $\gamma$, the planar penetration depth $\lambda_{ab}$, and the planar superconducting correlation length $\xi_{ab}$. $c_L$ is the Linderman number representing the fraction of the vortex separation bridged by thermal fluctuations required to melt the lattice. $G_i$ parameterizes the scale of thermal fluctuations, and thus the size of the fluctuating regime between $H_{\rm vs}(t)$ and $H_{\rm c2}(t)$, which becomes smaller with decreasing temperature. At zero temperature, $H_{\rm c2} (0)=H_{\rm vs}(0)$ in this model.  We performed least-squared fits of the vortex melting line by evaluating the above equation with $\alpha$ and $H_{\rm c2} (0)$ as free parameters. $H_{\rm c2} (0)$ determined in this manner for Y123~\cite{ramshaw12} and Tl$_2$Ba$_2$CuO$_{6+\delta}$ was reported to be consistent with that measured directly with thermal conductivity ~\cite{griss14}. Therefore, in the main text we refer to the critical field determined in this manner as the mean-field upper critical field $H_{\rm c2}^{\rm MF}$. 

A major source of uncertainty in our determination of $H_{\rm c2}^{\rm MF}$ is that it assumes a particular form of the vortex melting curve to the lowest temperature we are able to measure Hg1201 in pulsed magnetic fields, $\approx 1.5$~K. For Y123 with $p\sim0.11$, a surprisingly small value of $H_{\rm c2}^{\rm MF}(0)\approx 22~$T was found~\cite{griss14}. This result was apparently confirmed by recent NMR measurements in Y123~\cite{Zhou13148NMR}. However, thermal conductivity and torque magnetometry down to $0.3$~K  showed signs that the thermodynamic $H_{\rm c2}$ is significantly larger than the reported $H_{\rm c2}^{\rm MF}(0)$~\cite{yu16}. The value of $H_{\rm c2}$ in Y123 is still strongly debated. For Hg1201, the lowest $H_{\rm c2}^{\rm MF}(0)$ we have determined is $43\pm 5~{\rm T}$, which limits the ability to use sensitive thermal conductivity and NMR probes of superconductivity at low temperatures since the maximum DC magnetic field available is 45 tesla. The $H_{\rm c2}^{\rm MF}(0)$ in Fig.~\ref{PDTc}B should be taken as lower bounds on the critical field. It nevertheless reflects the doping dependence of superconductivity indicating maximized values upon approaching the Fermi-surface reconstruction phase boundaries.


\subsection*{Competing interest statement} The authors declare no competing interests.
}

\showmatmethods{} 

\acknow{We acknowledge L. Civale and S. Eley for support in performing DC susceptibility characterization of samples. The high-magnetic field measurements and sample preparation was supported by the US Department of Energy BES ‘Science of 100 tesla’ grant. The National High Magnetic Field Laboratory - PFF facility and related technical support is funded by the National Science Foundation Cooperative Agreement Number DMR-1644779, the State of Florida and the U.S. Department of Energy.  Crystal growth was supported by the LANL Laboratory Directed Research and Development (LDRD) program. 
}

\showacknow{} 

\bibliography{pnas-sample}

\newpage

\end{document}